\documentclass{eas}
\usepackage{graphicx}
\usepackage{natbib}

\usepackage{floatrow}
\usepackage{subfigure}
\usepackage{amsmath,amssymb}
\usepackage{mathrsfs}
\usepackage{mathabx}

\usepackage{txfonts}
\usepackage{pifont}

\usepackage{hyperref}

\hypersetup{
   colorlinks=true,
   citecolor=blue,
   linkcolor=blue,
   urlcolor=black
   }
\newcommand\apjl{The Astrophysical Journal Letters}
\newcommand\aap{Astronomy \& Astrophysics}

\newcommand{\bs}[1]{\boldsymbol{#1}}
\newcommand{\ve}[1]{\textbf{#1}}

\begin{document}

\title{Atmospheric tides and their consequences on the rotational dynamics of terrestrial planets}
\runningtitle{Atmospheric tides and their consequences on the rotation of planets}
\author{P. Auclair-Desrotour}\address{LAB, Universit\'e de Bordeaux, CNRS UMR 5804, Universit\'e de Bordeaux - B\^at. B18N, All\'ee Geoffroy Saint-Hilaire, CS50023, 33615 Pessac Cedex, France}
\author{J. Laskar}\address{IMCCE, Observatoire de Paris, CNRS UMR 8028, PSL, 77 Avenue Denfert-Rochereau, 75014 Paris, France}
\author{S. Mathis$^{3,}$}\address{Laboratoire AIM Paris-Saclay, CEA/DRF - CNRS - Universit\'e Paris Diderot, IRFU/SAp Centre de Saclay, F-91191 Gif-sur-Yvette Cedex, France}
\address{LESIA, Observatoire de Paris, PSL Research University, CNRS, Sorbonne Universit\'e, UPMC Univ. Paris~6, Univ. Paris Diderot, Sorbonne Paris Cité, 5 place Jules Janssen, F-92195 Meudon, France}

\begin{abstract}
Atmospheric tides can have a strong impact on the rotational dynamics of planets. They are of most importance for terrestrial planets located in the habitable zone of their host star, where their competition with solid tides is likely to drive the body towards non-synchronized rotation states of equilibrium, as observed in the case of Venus. Contrary to other planetary layers, the atmosphere is sensitive to both gravitational and thermal forcings, through a complex dynamical coupling involving the effects of Coriolis acceleration and characteristics of the atmospheric structure. These key physics are usually not taken into account in modelings used to compute the evolution of planetary systems, where tides are described with parametrised prescriptions. In this work, we present a new \textit{ab initio} modeling of atmospheric tides adapting the theory of the Earth's atmospheric tides \citep[][]{CL1970} to other terrestrial planets. We derive analytic expressions of the tidal torque, as a function of the tidal frequency and parameters characterizing the internal structure (e.g. the Brunt-V\"ais\"al\"a frequency, the radiative frequency, the pressure heigh scale). We show that stratification plays a key role, the tidal torque being strong in the case of convective atmospheres (i.e. with a neutral stratification) and weak in case of atmosphere convectively stable. In a second step, the model is used to determine the non-synchronized rotation states of equilibrium of Venus-like planets as functions of the physical parameters of the system. These results are detailed in \cite{ADLM2016a} and \cite{ADLM2016b}.
\end{abstract}
\maketitle
\section{Introduction}

Like solid and oceanic tides, atmospheric tides are responsible for internal variations of mass distribution and dissipation of energy, and thus contribute to the dynamical and rotational evolution of planetary systems \citep[e.g.][]{Kaula1964,Hut1981}. This contribution is often negligible compared to the two formers because of the weak energy of tides in the atmosphere relatively to the one in solid and oceanic layers. For example, the major part of the current total tidal dissipation of the Earth is due to the ocean. However, the atmospheric tidal dissipation can be strong in the case of Venus-like planets, i.e. planets basically composed of a solid core and a massive atmosphere, and rotating slowly. Indeed, as showed by early studies of the rotational dynamics \citep[cf. Venus,][]{GS1969,DI1980,CL01,CL2003,Correia2003}, the observed retrograde rotation rate of Venus is a state of equilibrium resulting from the competition between atmospheric and solid tides. Without the action of atmospheric tides, gravitational tides would torque Venus to spin-orbit synchronization.

The large number of new planetary systems discovered during the two past decades thanks to the CoRoT and \textit{Kepler} space missions \citep[e.g.][]{Baglin2006,Fabryckyb2012}, and particularly those hosting terrestrial planets in the habitable zone, has enhanced the importance of this effect of atmospheric tides on the rotational dynamics. Among the discovered bodies, configurations such as Venus' one are probable. As a consequence, atmospheric tides are likely to have a strong impact on the rotational dynamics of these planets. They can also affect indirectly their climate and atmospheric general circulation (e.g. differential rotation, tides induced flows, heating), which are of great importance regarding their surface habitability. For all these reasons, computing the long term orbital and rotational evolution of planetary systems and characterizing the atmosphere dynamics of the detected planets now requires to take into account the effects of atmospheric tides in a realistic way.

Yet, this is not the case of simplified modelings commonly used in celestial dynamics. Some of them, like the so-called \emph{Kaula's model} \citep[][]{Kaula1964}, set the dissipation to a constant, and thus do not describe the dependence of tidal dissipation on the tidal frequency, although this makes the dissipated energy vary over several orders of magnitude. Others include considerations on fluid dynamics based on a geophysical modeling \citep[][]{ID1978,DI1980,CL01} but have to introduce a parametrised regularization to prevent the dissipated energy to diverge at spin-orbit synchronization. By using a General Circulation Model (GCM) to compute thermal tides numerically, \cite{Leconte2015} recover this needed regular behaviour described by these modeling in the case of Venus with a physical approach. However, this method can be heavy to implement to explore the full domain of parameters, because of its possibly high computational cost. Moreover, it is important to understand in details the complex physical mechanisms involved in the atmospheric tidal dissipation. 

To address this question in the general case, we developed a new global \textit{ab initio} modeling based on the state of the art work of \cite{CL1970}, who treat the case of the Earth's atmospheric tides. In this model, the internal dissipative processes which are responsible for the behaviour of the tidal response at the vicinity of synchronization are taken into account using a \emph{Newtonian cooling} \citep[e.g.][]{LM1967}. We present here the outlines of the modeling and show how it can be applied to provide a prediction about the atmospheric structure and its impact on the rotational evolution of planets. For technical details, the reader may refer to \cite{ADLM2016a} and \cite{ADLM2016b}.



\begin{figure}[htb]
\centering
{\includegraphics[width=0.5\textwidth]{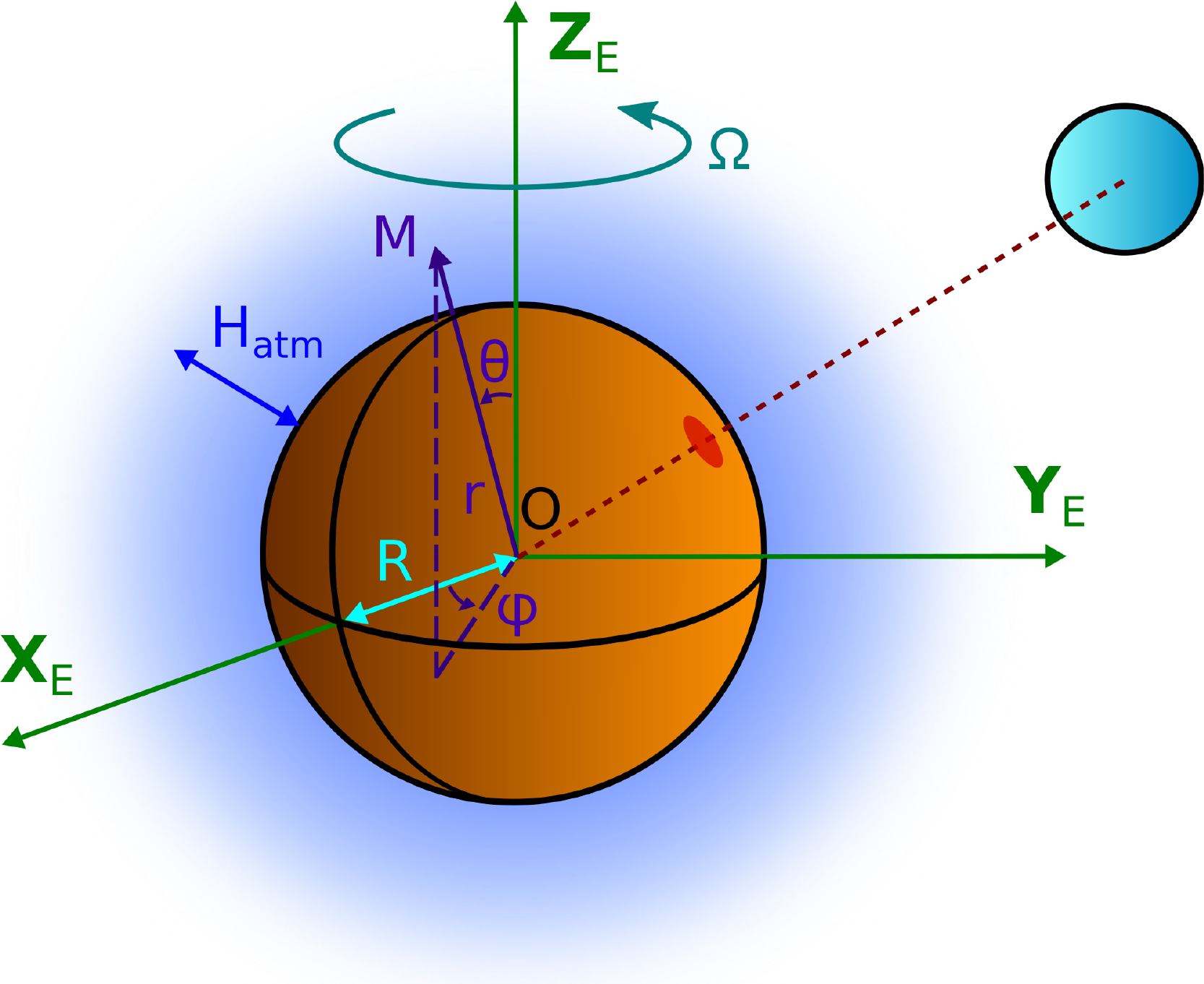}
\textsf{\caption{\label{fig:systeme}  System of spherical coordinates associated with the equatorial reference frame of the planet. Figure extracted from \cite{ADLM2016a}.} }}
\end{figure}

\section{Tidal waves dynamics}
\label{sec:dynamics}

We give here an overview of the dynamics of the modeling of the atmospheric tidal response. The results presented in this section are detailed in \cite{ADLM2016a}. We consider a terrestrial planet of radius $ R $ rotating at the angular velocity $ \Omega $ (the corresponding spin vector being $ \bs{\Omega} $), and tidally excited by the thermal and gravitational forcings of the host star (Fig.~\ref{fig:systeme}). To study the tidal response of the planet's atmosphere, we introduce the equatorial frame co-rotating with the planet $ \mathcal{R}_{\rm E}: \left\{ \ve{X}_{\rm E} , \ve{Y}_{\rm E}, \ve{Z}_{\rm E} \right\} $, where $ \ve{X}_{\rm E} $ and $ \ve{Y}_{\rm E} $ define the equatorial plane and $ \ve{Z}_{\rm E} = \bs{\Omega} / \left| \Omega \right| $ is aligned with the spin axis of the planet. In this frame, the position of a point $ M $ is described by the usual system of spherical coordinates $ \left( r , \theta , \varphi \right) $ and the associated basis $ \left( \ve{e}_r , \ve{e}_\theta , \ve{e}_\varphi \right) $, $ r $ being the radius, $ \theta $ the colatitude and $ \varphi $ the longitude ($ \ve{r} = r \, \ve{e}_r $ stands for the corresponding position vector); the time is denoted $ t $. The physics of the atmosphere are represented by the pressure ($p$), density ($ \rho$), temperature ($ T $), velocity ($ \ve{V} $) and gravity ($g$) distributions, a given quantity $ f $ being written as $  f \left( \ve{r} , t \right) = f_0 \left( \ve{r} \right) + \delta f \left( \ve{r} , t \right) $, where the superscript $ _0 $ refers to background distributions and $ \delta $ to the tidal perturbation. The analytic treatment of the dynamics is simplified by assuming the approximations listed below:
\begin{enumerate}
\item \emph{Solid body rotation}: the atmosphere is supposed to rotate uniformly with the planet, so that mean flows are ignored ($ \ve{V}_0 = \ve{0}$).
\item \emph{Moderate rotation}: the centrifugal acceleration due to the rotation of the body is negligible compared to its self-gravity. This implies that $ \Omega \ll \Omega_{\rm c} $, the parameter $ \Omega_{\rm c} = \sqrt{g / r} $ being the so-called \emph{critical Keplerian angular velocity}. 
\item \emph{Spherical symmetry of the background}: the background distributions of gravity, pressure, density and temperature are assumed to vary along the radial direction only. Thus, the horizontal variations of the atmospheric structure are neglected.
\item \emph{Perfect gas approximation}: the fluid is supposed to follow the law of perfect gas and to be characterized by the constant adiabatic exponent $ \Gamma_1 = 1.4 $. We also introduce the parameter $ \kappa = \left( \Gamma_1 - 1 \right) / \Gamma_1 $.
\item \emph{Uniform composition}: the atmosphere is uniform in composition, this later being characterized by the specific gas constant $ \mathscr{R}_{\rm s} = \mathscr{R}_{\rm GP} / M $ (where $ \mathscr{R}_{\rm GP} $ designates the perfect gas constant and $ M $ the molecular weight of the atmosphere). The thermal capacity per unit mass of the atmosphere is defined by $ C_{\rm p} = \mathscr{R}_{\rm s} / \kappa $.
\item \emph{Cowling approximation} \citep[][]{Cowling1941}: the self-gravitational perturbation resulting from the tidal variations of mass distribution is not taken into account. 
\item \emph{Traditional approximation}: the latitudinal components of Coriolis accelerations ($ 2 \Omega \sin \theta V_r $ and $ 2 \Omega \sin \theta V_\varphi $) are ignored. The conditions of validity of this approximation are discussed in details in \cite{Mathis2009} and \cite{ADLM2016a}.
\item \emph{Linear approximation}: the perturbation is assumed to be small enough to neglect non-linear couplings, i.e. $ \delta f / f_0 \ll 1 $. 
\item \emph{Newtonian cooling}: the zero--order effect of radiative cooling and thermal diffusion is described with the sink power per unit mass $ J_{\rm NC} =   \sigma_0 C_{\rm p} \delta T $, where the radiative frequency $ \sigma_0 $ corresponds to the inverse of the effective radiative time of the atmosphere. 
\end{enumerate}

\noindent The compressibility of the atmosphere is characterized by the sound frequency $ c_s $ and its stratification by the Brunt-V\"ais\"al\"a frequency $ N $, defined by 

\begin{equation}
N^2 = g \left[ \frac{1}{\Gamma_1} \frac{d \ln p_0}{dr} - \frac{d \ln \rho_0}{dr} \right].
\end{equation}

\begin{figure}[htb]
\centering
{\includegraphics[width=0.98\textwidth]{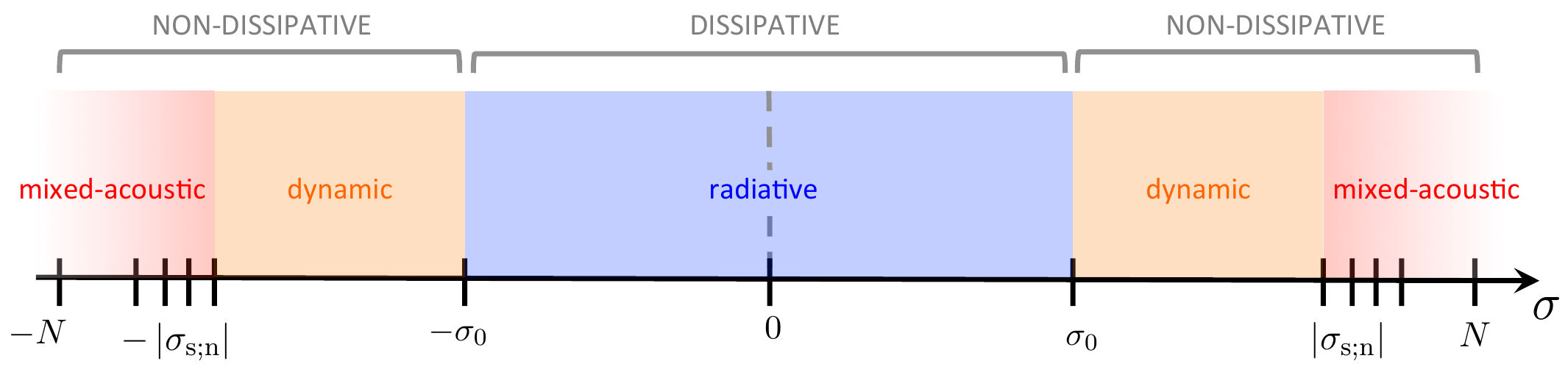}
\textsf{\caption{\label{fig:spectrum_waves} Frequency spectrum of the atmospheric tidal response. The identified regimes are given as functions of the tidal frequency. Figure extracted from \cite{ADLM2016a}.} }}
\end{figure}

The tidal perturbation being periodic in time and longitude, the gravitational potential $U$ and the thermal power per unit mass $ J $ of the forcing, as well as the perturbed quantities are expanded in Fourier series. Thanks to the traditional approximation, Fourier coefficients can be themselves expanded in series of functions with separated coordinates. Hence, a fluctuation $ \delta f $ of any quantity writes 

\begin{equation}
\delta f = \sum_{n,m,\sigma} \delta f_n^{m,\sigma} \left( r \right) \Theta_n^{m,\nu} \left( \theta \right) e^{i \left( \sigma t + m \varphi \right)},
\end{equation} 

\noindent where we have introduced the tidal frequency $ \sigma $,  the longitudinal and latitudinal wavenumbers, $ m $ and $ n $ respectively, the spin parameter $ \nu = 2 \Omega / \sigma $, the Hough functions $ \Theta_n^{m,\nu} $ \citep[see][]{Hough1898,LS1997,ADLM2016a,Wang2016} and the corresponding vertical profiles $ \delta f_n^{m,\sigma} $. We will not develop here technical aspects concerning solutions $ \delta f_n^{m,\sigma} $, which are described in \cite{ADLM2016a}. However, the physical setup allows us to identify the possible regimes for the atmospheric tidal response (Fig.~\ref{fig:spectrum_waves}):
\begin{enumerate}
 \item \emph{Mixed-acoustic regime} for tidal frequencies greater than the \emph{Lamb frequencies} $ \sigma_{s ; n} = \sqrt{\Lambda_n^{m,\nu}} c_s / R $ of the gravest acoustic modes ($ \Lambda_n^{m,\nu} $ designates the eigenvalue associated to the Hough function $ \Theta_n^{m,\nu} $). 
 \item \emph{Dynamic regime} for $ \sigma_0 \ll \left| \sigma \right| \ll \left| \sigma_{\rm s ; n} \right| $. In this regime, which corresponds to the Earth's semi-diurnal tide typically \citep[][]{CL1970}, the effect of dissipation on the structure of tidal waves can be ignored. 
 \item \emph{Radiative regime} for low tidal frequencies ($ \left| \sigma \right| \lesssim \sigma_0 $). This regime corresponds to the vicinity of synchronization. The atmospheric tidal response is regular thanks to dissipative processes, which damp tidal waves. 
\end{enumerate}

\begin{figure}[htb]
\centering
{\includegraphics[width=0.98\textwidth]{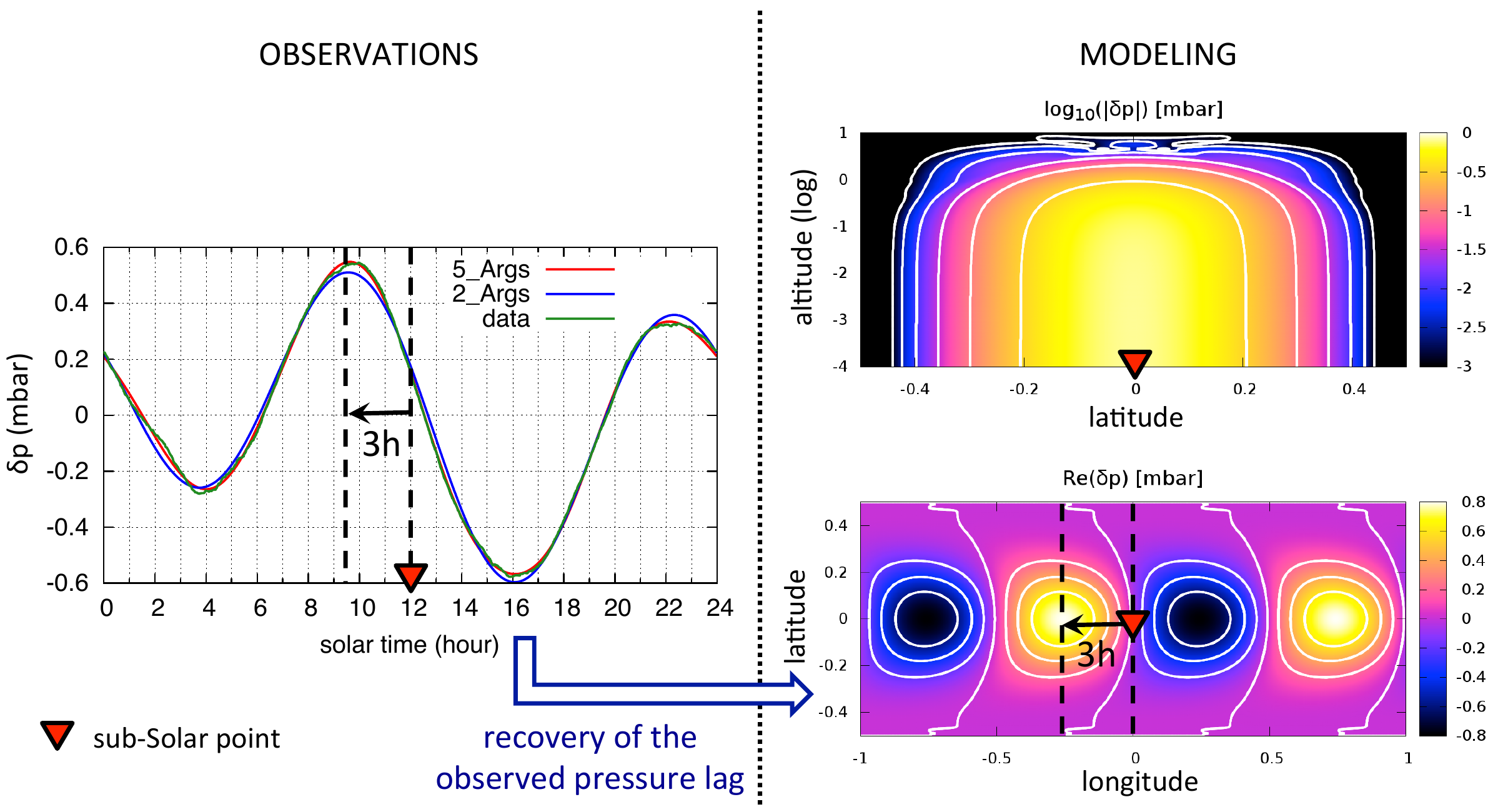}
\textsf{\caption{\label{fig:schema_pression} {\it Left~-} Daily averaged Solar semidiurnal surface barometric variations (mbar) measured at Fontainebleau (lat. 48,363$^\degree$ N) over the full year 2013 by J. Laskar. The time is given in hours. {\it Right~- } Earth Solar semidiurnal atmospheric tidal response computed with the model: amplitude of pressure variations as a function of latitude and altitude (top) and surface pressure variations as a function of longitude and latitude (bottom). } }}
\end{figure}

\noindent As showed by Fig.~\ref{fig:schema_pression}, we recover with this modeling the observed lag and amplitude of the Earth's semidiurnal surface pressure oscillations (i.e. oscillations corresponding to the tidal frequency $ \sigma = 2 \left( \Omega - n_{\rm orb} \right) $, where $ n_{\rm orb} $ designates the orbital frequency of the Earth). Such oscillations are mainly due to the first quadrupolar gravity mode ($m=2$, $ n = 0$).

\section{Tidal torque: the key role played by stratification}
\label{sec:torque}

\begin{figure}[htb]
\centering
{
\includegraphics[width=0.8\textwidth]{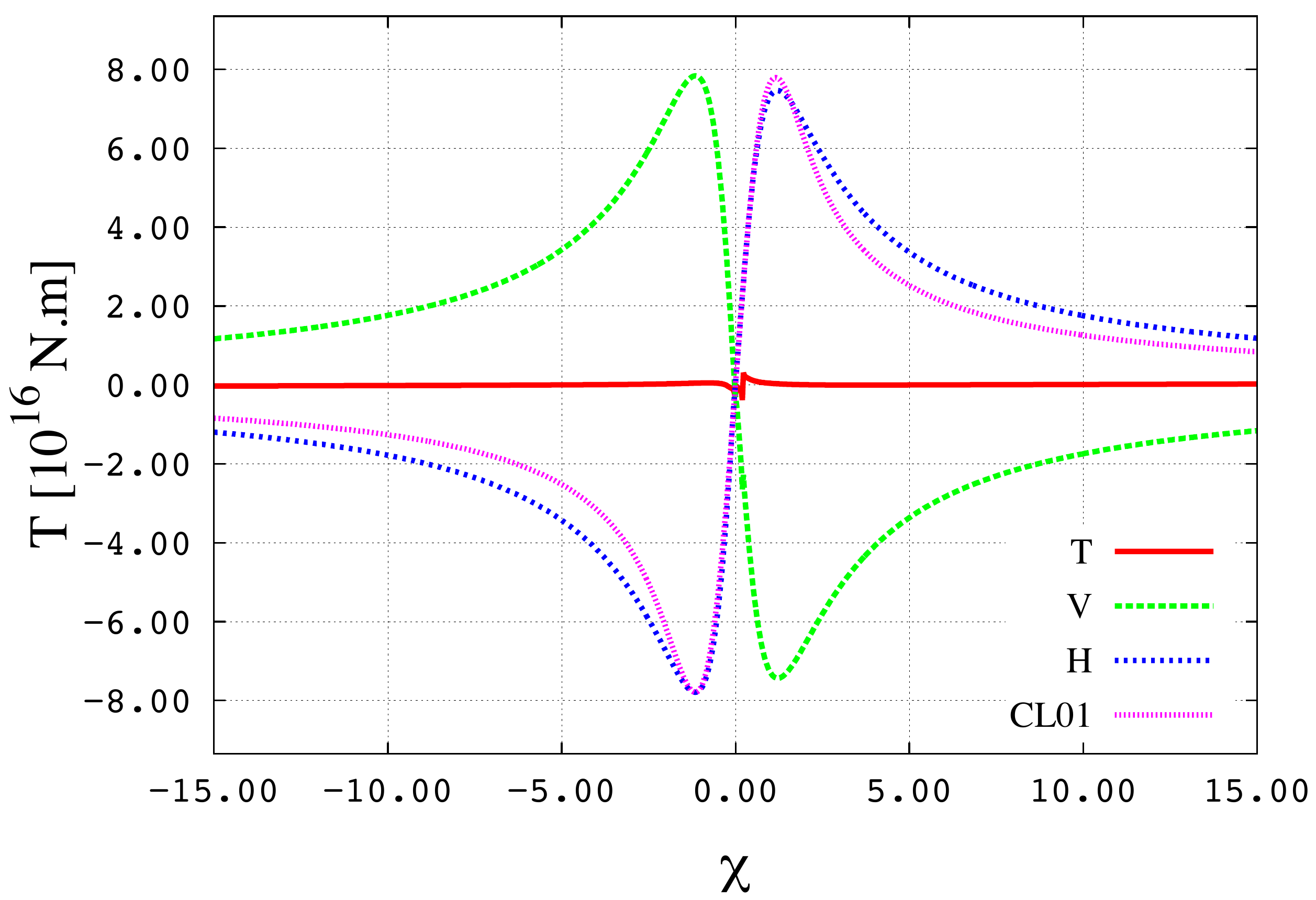}
\includegraphics[width=0.98\textwidth]{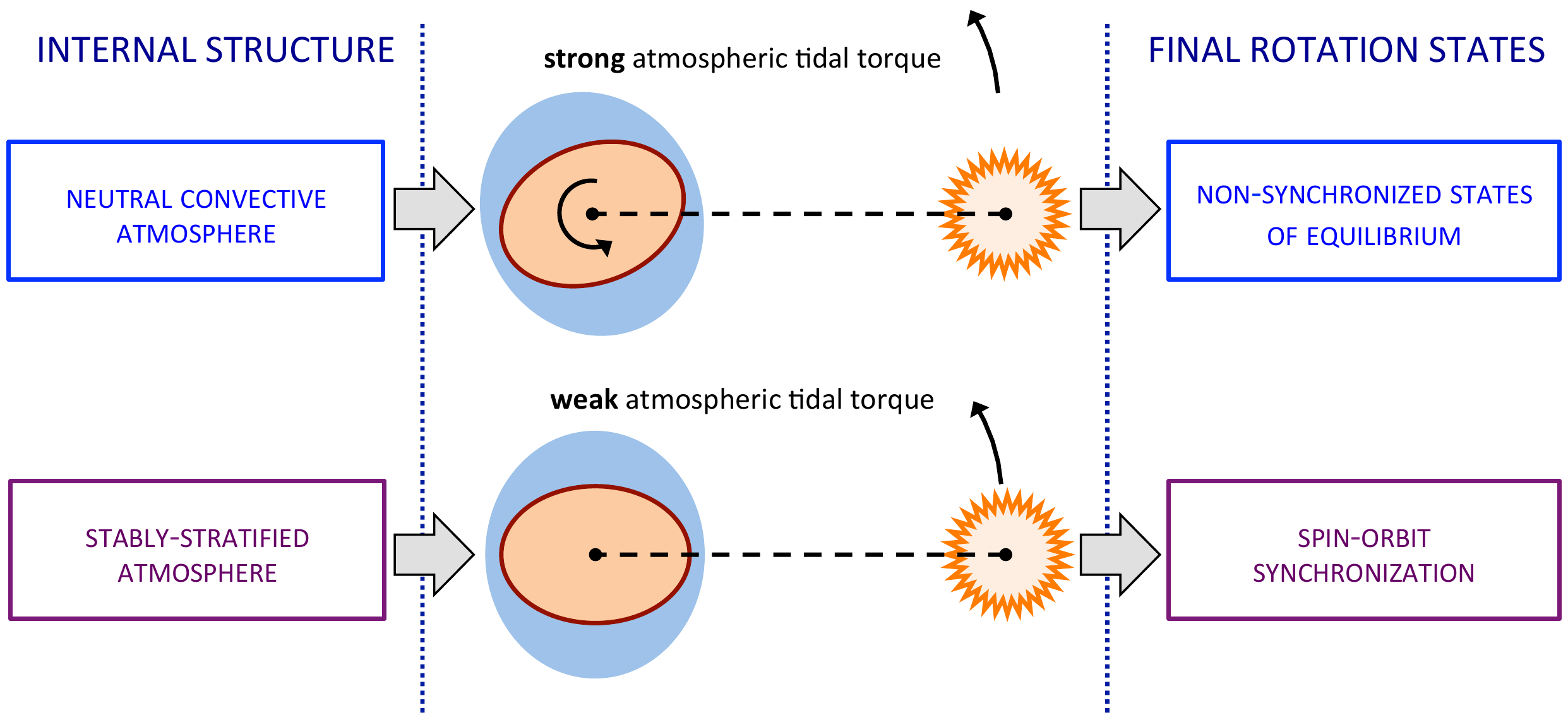}
\textsf{\caption{\label{fig:Venus_torque} {\it Top~-} Tidal torque exerted on the stably-stratified atmosphere of a Venus-like planet and due to the quadrupolar component of the Solar semidiurnal tide. The total torque (T, red continuous line) and its \emph{horizontal} (H, blue dashed line) and \emph{vertical} (V, green dashed line) components, as well as the early model of \cite{CL01} (CL01, pink dashed line), are plotted as functions of the normalized frequency of the perturber $ \chi = \left( \Omega - n_{\rm orb} \right) / n_{\rm orb} $. In the case of a convective atmosphere, the total torques reduces to the H-component. {\it Bottom~-} Predictive aspects related to the dependence of the atmospheric tidal torque on the stability of stratification. Figure extracted from \cite{ADLM2016a}.} }}
\end{figure}

The rotational evolution of the planet is affected by the tidal torque exerted on the atmosphere with respect to the spin axis of the planet, denoted $ \mathcal{T}^{2,\sigma} $ in the case of a quadrupolar perturbation. This torque is deduced straightforwardly from the variations of mass distribution \citep[see the expression given by][p.~319]{Zahn1966a}. Hence, for a slowly rotating planet with a convective atmosphere, one obtains \citep[][]{ADLM2016a}

\begin{equation}
\mathcal{T}_{\rm neutral}^{2,\sigma} = 2 \pi R^2 \frac{\kappa \rho_{\rm s}}{g}  U_2 J_2 \frac{\sigma}{\sigma^2 + \sigma_0^2},
\label{torque_neutral}
\end{equation}

\noindent where $ \rho_{\rm s} $ designates the surface density of the atmospheric layer, and $ U_2 $ and $ J_2 $ the quadrupolar components of the gravitational tidal potential and thermal forcing respectively. This torque, plotted on Fig.~\ref{fig:Venus_torque} in the case of a Venus-like planet (blue dashed line), is similar to those obtained by \cite{Leconte2015} with numerical simulations using GCMs and early parametrised modelings \citep[e.g.][pink dashed line]{CL01}. It corresponds to the \emph{horizontal} (H) hydrostatic adjustment of the atmosphere with a lag depending on its thermal inertia and dissipative properties. 

The expression of the torque in the case of a stably stratified isothermal atmosphere ($ N^2 \gg \left| \sigma \right| $) is given by \citep[see][Eq.~141]{ADLM2016a}

\begin{equation}
\mathcal{T}_{\rm strat}^{2,\sigma} = - 2 \pi R^2 H U_2 \sum_{n} C_{2,n,2}^{2,\nu} \Im \left\{  \mathcal{A}_n J_2 + \mathcal{B}_n U_2  \right\},
\end{equation}

\noindent where $ \Im $ designates the imaginary part of a complex number, $ H = p_0 / \left( g \rho_0 \right) $ the pressure height scale of the atmosphere, the $ C_{2,n,2}^{2,\nu} $ the projection coefficients quantifying the effect of the Coriolis acceleration on the structure of the perturbation, and $ \eta_{s ; n} $, $ \mathcal{A}_n $ and $ \mathcal{B}_n $ parameters depending on the tidal frequency and vertical wavenumber of the $ \left( n , 2 \right) $-mode. In the vicinity of synchronization, the Archimedean force inhibits tidal motions in the vertical direction. The integral of the resulting variations over the air column vanishes when  $ \sigma \rightarrow 0 $. As a consequence, the tidal torque (Fig.~\ref{fig:Venus_torque}, red continuous line), which receives the contribution of \emph{vertical} (V) displacements (green dashed line), is weak compared to $\mathcal{T}_{\rm neutral}^{2,\sigma}$ (H, blue dashed line). Considering planetary rotational evolution, the planet is thus likely to be driven towards non-synchronized states of equilibrium in the convective case and towards spin-orbit synchronization in the stably-stratified one, as illustrated by Fig.~\ref{fig:Venus_torque}.

\section{Consequence on the equilibrium rotation of Venus-like planets}
\label{sec:equilibrium}

\begin{figure}[htb]
\centering
{\includegraphics[width=0.47\textwidth]{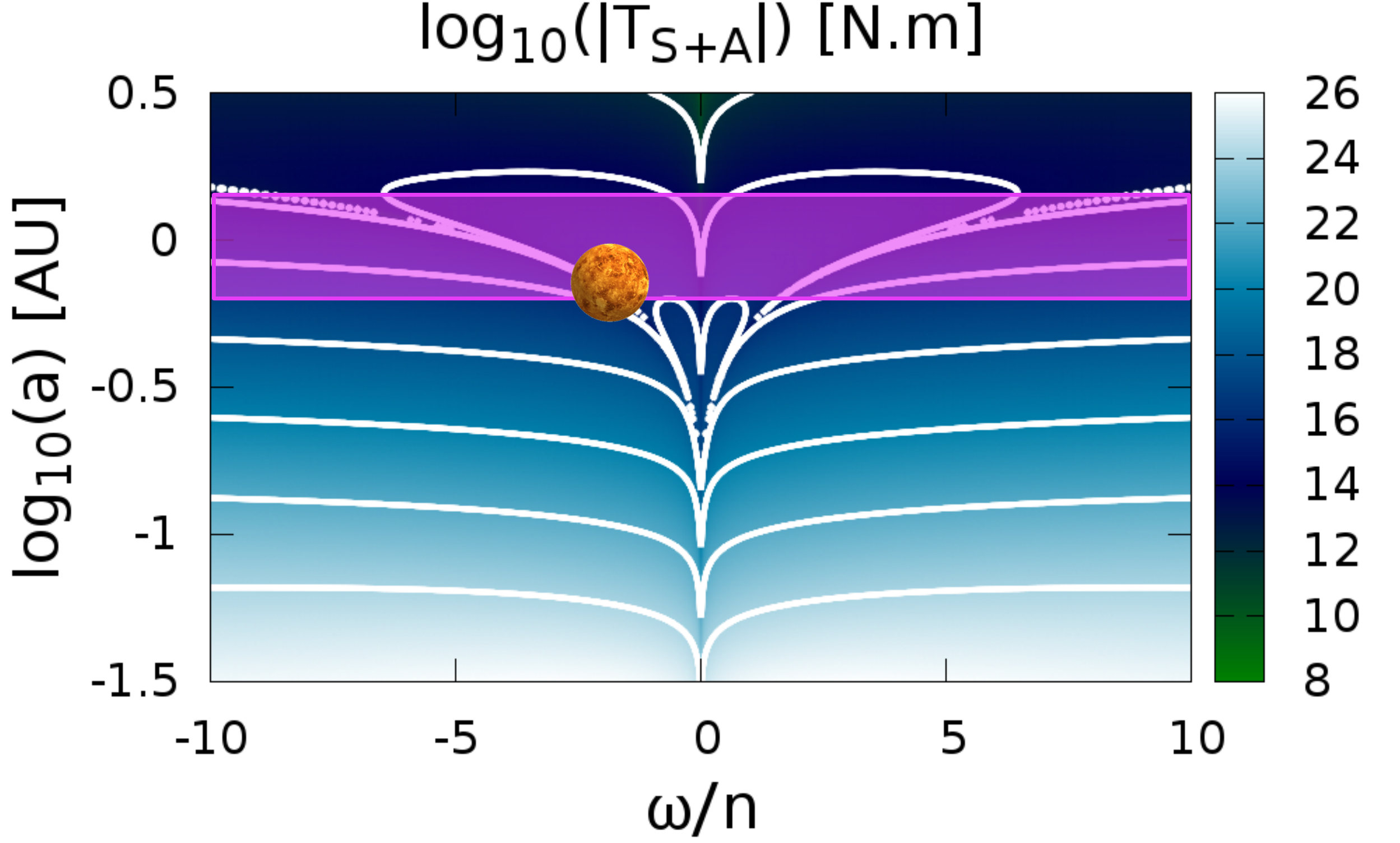}
\hspace{0.5cm}
\includegraphics[width=0.47\textwidth]{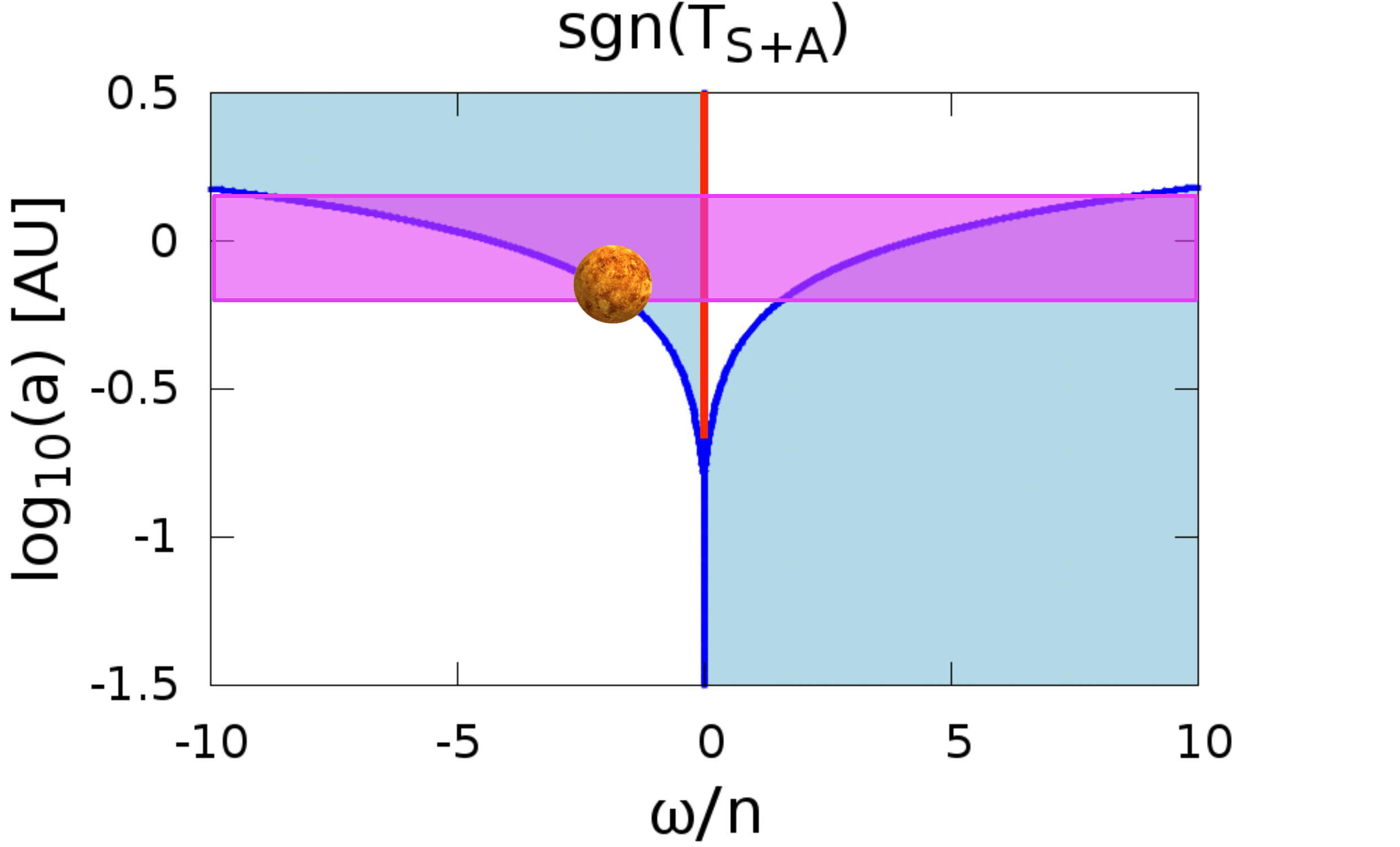} 
\textsf{\caption{\label{fig:Venus_equilibrium} Tidal torque exerted on a Venus-like planet by the host star of a planetary system as function of the difference to spin-orbit synchronization (horizontal axis) and logarithmic star planet distance (vertical axis). The planet is composed of a solid part and a convective atmosphere. {\it Left~-} Tidal torque in logarithmic scale. {\it Right~-} Sign of the torque. A positive (negative) torque corresponds to white (blue) regions. Stable (unstable) state of equilibrium are represented by  blue (red) lines. Figure extracted from \cite{ADLM2016b}.} }}
\end{figure}

By using the torque computed in the case of a convective atmosphere and the so-called \emph{Maxwell rheology} to describe the tidal response of the solid part, we derive the final rotation states of equilibrium of a Venus-like planet as functions of the physical parameters of the system \citep[see][for details]{ADLM2016b}. In this framework, the stability of the states of equilibrium appears to depend on the hierarchy of frequencies associated to dissipative processes, i.e. $ \sigma_0 $ for the atmosphere, and the Maxwell relaxation frequency of the material $ \sigma_{\rm M} $ for the solid part. These results are illustrated by Fig.~\ref{fig:Venus_equilibrium}, where the total torque exerted on the planet and its sign are plotted as functions of the difference to synchronization and distance to the host star. Note that the Maxwell model underestimates the tidal torque of the solid part for $ \left| \sigma \right| \gg \sigma_{\rm M} $, which can be detrimental to a quantitative prediction. In order to address this point, the Maxwell rheology could be replaced by the Andrade rheology \citep[][]{Andrade1910}, this later giving a better description of the solid tidal response. Four non-synchronized states instead of two would thus be obtained, as demonstrated by \cite{Leconte2015}.


\section{Conclusions}
\label{sec:conclusions}

In this work, we developed an \textit{ab initio} modeling for the atmospheric tides of terrestrial planets and applied it to the case of Venus-like planets. We established the dependence of the tidal response on the parameters of the internal structure, and particularly the stability of the vertical stratification. If this later is convective, we recover the tidal torque given by GCMs and early modelings. Otherwise, stable stratification decreases the amplitude of the torque of several orders of magnitude, which leaves the planet evolve towards spin-orbit synchronization. The chosen analytic treatment provides both a diagnostic of the physics involved in atmospheric tides and a predictive tool to explore the full domain of possible parameters. It will be improved in the future with the introduction of general circulation, means flows being able to modify significantly the structure of tidal waves \citep[see][]{Volland1974a}.

\begin{acknowledgements}
P. Auclair-Desrotour and S. Mathis acknowledge funding by the European Research Council through ERC grants WHIPLASH 679030 and SPIRE 647383. This work was also supported by the Programme National de Plan\'etologie (CNRS/INSU) and CoRoT/Kepler and PLATO CNES grant at CEA-Saclay. 
\end{acknowledgements}


\bibliographystyle{aa}  
\bibliography{astrofluid2016} 

\end{document}